\documentstyle[aps,epsf]{revtex}
\begin{document}

\title{Excitation energies from density functional perturbation theory}
\author{Claudia Filippi}
\address{Department of Physics, University of Illinois at Urbana-Champaign,
Urbana, Illinois 61801}
\author{C. J. Umrigar}
\address{Cornell Theory Center and Laboratory of Atomic and Solid State
Physics, Cornell University, Ithaca, New York 14853}
\author{Xavier Gonze}
\address{Unit\'e de Physico-Chemie et de Physique des Mat\'eriaux,
Universit\'e Catholique de Louvain, B-1348 Louvain-la-Neuve, Belgium}
\maketitle
\begin{abstract}
We consider two perturbative schemes to calculate excitation energies,
each employing the Kohn-Sham Hamiltonian as the unperturbed system.
Using accurate exchange-correlation potentials generated from essentially
exact densities and their exchange components determined by a recently
proposed method, we evaluate energy differences between the ground state
and excited states in first-order perturbation theory for the Helium,
ionized Lithium and Beryllium atoms.
It was recently observed that the zeroth-order excitations energies, simply
given by the difference of the Kohn-Sham eigenvalues, almost always lie between
the singlet and triplet experimental excitations energies, corrected for
relativistic and finite nuclear mass effects.
The first-order corrections provide about a factor of two improvement
in one of the perturbative schemes but not in the other.
The excitation energies within perturbation theory are
compared to the excitations obtained within $\Delta$SCF and time-dependent
density functional theory.
We also calculate the excitation energies in perturbation theory
using approximate functionals such as the local density approximation and
the optimized effective potential method with and without the Colle-Salvetti
correlation contribution.
\end{abstract}

\pacs{PACS numbers: 31.15.Ew, 31.50.+w, 31.15.Md, 71.10.-w}
%31.15.Ew Density-functional theory
%31.50.+w Excited states
%31.15.Md Perturbation theory
%71.10.-w Theories and models of many electron systems

\section{Introduction}
\label{s1}
Density functional theory is usually employed to study ground states.
However, it is clear that density functional theory can in principle be used
to calculate excitation energies since, according to the Hohenberg-Kohn
theorem~\cite{HK}, the density determines the external potential (aside from
an additive constant) and, consequently, the entire spectrum.
In principle, one could construct functionals of the ground state densities
that yield excitation energies.

There are several proposals in the literature to deal with excited states.
Gunnarson and Lundquist~\cite{GL} pointed out that the original Hohenberg-Kohn
theorems can be generalized to apply to the lowest energy state of any given
symmetry.
Von Barth~\cite{VB} extended density functional theory to states of mixed
symmetry following the observation of Ziegler {\it et al.}~\cite{ZRB} that
the X$\alpha$ method is designed to describe states that have single
determinant wave functions in the absence of interaction.
The practical recipe consists of calculating, within the local density
approximation, the energies of states of mixed symmetry represented by a single
determinant and linearly combining them to construct the energies of the states
of pure symmetry~\cite{VB,ZRB,GJ}.
Further developments and applications of this procedure, known as $\Delta$SCF,
can be found in Refs.~\cite{D,DGW,GWBDG,GRC}.
Another generalization of density functional theory to treat excited
states is the ensemble density functional theory~\cite{T,GOK,N}.
%but little knowledge is presently available about this excited-state
%exchange-correlation functional and its dependence on the ensembles.
Recently, a different and quite promising method to calculate excited
states based on time-dependent density functional theory has been
developed~\cite{C,CCS,PGG1,PG,PGG2}.

An alternative approach~\cite{G} is to calculate excitation energies
perturbatively using the non-interacting Kohn-Sham Hamiltonian
as the zeroth-order Hamiltonian.
In the Kohn-Sham formulation of density functional theory~\cite{KS}, the
interacting electron system is replaced by a system of non-interacting electrons
in an effective local potential that yields the true ground state density.
In the perturbative approach, the zeroth-order
excitation energies are simply the differences of Kohn-Sham eigenvalues.
Remarkably, it has been recently shown~\cite{SUG} that, for atomic systems, the
excitation energies obtained as differences of Kohn-Sham eigenvalues
lie between the triplet and singlet experimental excitation
energies, corrected for relativistic and finite nuclear mass effects, with
the exception of the $1s\rightarrow 3d$ transition of ionized Lithium.
This finding is surprising since, except for the highest occupied orbital,
there is no strict relationship between the Kohn-Sham eigenvalues
and excitation energies.

In an attempt to improve on these results and determine the singlet-triplet
splitting, we calculate excitation energies to first-order in perturbation
theory.  We employ two variants of perturbation theory.
Interestingly, we find that the first-order corrections from the more obvious
{\it standard perturbation theory} make the excitation energies worse on
average while those from the {\it coupling-constant perturbation theory}
of G\"orling and Levy~\cite{GL1} yield about a factor of two improvement.
To perform these calculations, we employ excellent approximations to the
true Kohn-Sham exchange-correlation potentials determined by generating
an accurate density for the system of interest and then computing an
exchange-correlation potential that yields the desired density as the ground
state solution for the fictitious non-interacting system~\cite{proc,UG,review}.
Evaluating the first-order corrections also requires the knowledge of
the correct exchange component of the exchange-correlation potential as
determined in our previous work~\cite{separation}.
Standard perturbation theory has also been recently applied to calculate
ground state energies and the perturbation series based on various approximate
exchange-correlation functionals were found to be divergent for the systems
tested~\cite{W}.

In Sec.~\ref{s1}, we briefly introduce density functional theory and its
Kohn-Sham formulation. In Sec.~\ref{s2}, we present the coupling-constant
scheme and the formulae used to determine the excitation energies to first-
order in the coupling-constant. In Sec.~\ref{s3}, we give the results and
a comparison with other methods to evaluate excitation energies. The
perturbation theory scheme is also implemented using approximate
exchange-correlation functionals.
In Appendix~\ref{a1}, we derive the formulae for the first-order corrections
to the excitation energies.
%to the excitation energies and, in Appendix~\ref{a2}, we rewrite these
%expressions for the special case of closed shell systems.

\section{Theoretical background}
\label{s2}

In the Kohn-Sham formulation of density functional theory~\cite{KS}, the ground
state density is written in terms of single-particle orbitals obeying the
equations in atomic units ($\hbar=e=m=1$):
\begin{eqnarray}
\left\{-\frac{1}{2}\nabla^2+v_{\rm ext}({\bf r})+v_{\rm H}({\bf r})
+v_{\rm xc}\left(\left[\rho\right];{\bf r}\right)\right\}\psi_i
=\epsilon_i\psi_i,
\label{KS}
\end{eqnarray}
where
\begin{eqnarray}
\rho({\bf r})=\sum_{i=1}^N\left|\psi_i({\bf r})\right|^2.
\label{rho}
\end{eqnarray}
The electronic density is constructed by summing over the {\it N} lowest
energy orbitals where {\it N} is the number of electrons.
$v_{\rm ext}({\bf r})$ is the external potential and $v_{\rm H}({\bf r})$
the Hartree potential.
% defined as
%\begin{eqnarray}
%v_{\rm H}({\bf r})=
%\int \frac{\rho({\bf r}')} {\left|{\bf r}-{\bf r}'\right|}{\rm d}{\bf r}'.
%\label{hartree}
%\end{eqnarray}
The exchange-correlation potential $v_{\rm xc}\left(\left[\rho\right];
{\bf r}\right)$ is the functional derivative of the exchange-correlation
energy $E_{\rm xc}\left[\rho\right]$ that enters in the expression for
the total energy of the system:
\begin{eqnarray}
E=-\frac{1}{2}
\sum_{i=1}^N\int\psi_i\nabla^2\psi_i\,{\rm d}{\bf r}
+\int\,\rho\left({\bf r}\right)v_{\rm ext}
\left({\bf r}\right)\,{\rm d}{\bf r}
+\frac{1}{2}\int\!\!\int\frac{\rho({\bf r})
\rho({\bf r}')}{\left|{\bf r}-{\bf r}'\right|} {\rm d}{\bf r}\,{\rm d}{\bf r}'
+E_{\rm xc}\left[\rho\right].
\label{eq0}
\end{eqnarray}
The exchange-correlation functional is written as the sum of two separate
contributions for exchange and correlation,
\begin{eqnarray}
E_{\rm xc}\left[\rho\right]=E_{\rm x}\left[\rho\right]+
E_{\rm c}\left[\rho\right],
\end{eqnarray}
yielding a corresponding splitting in the exchange-correlation potential,
$v_{\rm xc}([\rho];{\bf r})=v_{\rm x}([\rho];{\bf r})+v_{\rm c}([\rho];{\bf r})$.
The definition of the exchange energy is in terms of the non-interacting
wave function $\Phi_0$, the Slater determinant constructed from the Kohn-Sham
orbitals, as
\begin{eqnarray}
{\rm E}_{\rm x}\left[\rho\right]=
\left<\Phi_0\right|V_{\rm ee}\left|\,\Phi_0\right>-
\frac{1}{2}\int\!\!\int\frac{\rho({\bf r})
\rho({\bf r}')}{\left|{\bf r}-{\bf r}'\right|}
{\rm d}{\bf r}\,{\rm d}{\bf r}',\label{enx0}
\end{eqnarray}
where $V_{\rm ee}$ is the electron-electron interaction.

\section{Perturbation theory}
\label{s3}

In many-body perturbation theory the Hamiltonian ${\cal H}$ is written as the
sum of a zeroth-order Hamiltonian ${\cal H}_0$ and a perturbing Hamiltonian
${\cal H}_1={\cal H}-{\cal H}_0$.

In the commonly used M\"oller-Plesset perturbation theory~\cite{MP},
the zeroth-order Hamiltonian is the Fock operator:
\begin{eqnarray}
{\cal H}_0^{\rm HF} = T + V_{\rm ext} + V_{\rm H} + V_{\rm x}^{\rm HF},
\end{eqnarray}
while in density functional perturbation theory
\begin{eqnarray}
{\cal H}_0 = T + V_{\rm ext} + V_{\rm H} + V_{\rm xc},\label{H0}
\end{eqnarray}
where $T$ is the kinetic operator and $V_{\rm x}^{\rm HF}$ is the non-local Hartree-Fock
exchange operator. Here and elsewhere, we employ the notation that potentials are
capitalized when summed over all the electrons.
The perturbation corresponding to the zeroth-order (Eq.~\ref{H0}) is clearly given by
\begin{eqnarray}
{\cal H}_1= V_{\rm ee}-V_{\rm H}-V_{\rm xc}.\label{H1}
\end{eqnarray}

The other variant of perturbation theory we consider is due to G\"orling and
Levy~\cite{GL1} and is based on the adiabatic connection Hamiltonian which
links the Kohn-Sham Hamiltonian with the fully interacting Hamiltonian keeping
the ground state density constant independent of $\alpha$:
\begin{eqnarray}
{\cal H}^\alpha=T+V_{\rm ext}+V_{\rm H}+V_{\rm xc}
+\alpha\left(V_{\rm ee}-V_{\rm x}-V_{\rm H}\right)
-V_{\rm c}^\alpha.\label{Halpha}
\end{eqnarray}
$v_{\rm c}^\alpha$ is second order in $\alpha$~\cite{GL1} and equals the correlation
potential when $\alpha=1$.
The zeroth-order Hamiltonian is again the Kohn-Sham Hamiltonian and the
perturbing Hamiltonian contains a term
\begin{eqnarray}
{\cal H}^{(1)}_1=\alpha \left(V_{\rm ee}-V_{\rm H}-V_{\rm x}\right),\label{H11}
\end{eqnarray}
which is linear in the perturbation parameter $\alpha$ and a component
$-v_{\rm c}^\alpha$ which contains second and higher order contributions~\cite{GL1}.
To any order in perturbation theory, the density equals the true ground state
density to that order in the coupling-constant~\cite{GL1}.

Let us consider the excitation of an electron from the space-spin orbital
$\phi_k$ to $\phi_\nu$.
In both perturbative schemes, the zeroth-order energy difference
equals the difference in the expectation value of the Kohn-Sham Hamiltonian
on the eigenstates $\Phi$ and $\Phi_{k\nu}$, obtained from $\Phi$ by exciting
an electron from orbital $k$ to orbital $\nu$, and is simply the
difference of the Kohn-Sham eigenvalues:
\begin{eqnarray}
\Delta{\rm E}^{(0)}=\epsilon_\nu-\epsilon_k.~\label{de0}
\end{eqnarray}
The correction to first-order in the coupling-constant (Eq.~\ref{H11})
is given by
\begin{eqnarray}
\Delta{\rm E}^{(1)}=
\left<\Phi_{k\nu}\left|V_{\rm ee}-V_{\rm H}-V_{\rm x}\right|\Phi_{k\nu}\right>-
\left<\Phi\left|V_{\rm ee}-V_{\rm H}-V_{\rm x}\right|\Phi\right>.\label{de1}
\end{eqnarray}
This expression simplifies for the Helium, ionized Lithium and Beryllium atoms,
that we use as examples in this work, all having ground states of $^1$S
symmetry.  The excited states are either singlets or triplets.
Let $\Phi_0$ be the Kohn-Sham ground state and $\Phi_{k\nu}$ the excited
state obtained by exciting an electron from a (doubly occupied) orbital $k$
to orbital $\nu$.
In this case, as shown in Ref.~\cite{G} and Appendix~\ref{a1}, the first-order
energy difference between the ground state and the state
$\Phi_{k\nu}$ depends on whether $\Phi_{k\nu}$ is a triplet or a singlet and is given
respectively by
\begin{eqnarray}
\Delta{\rm E}^{(1)}\left({\rm T},k\rightarrow\nu\right)=
\left<\phi_\nu\left|\hat{v}_{\rm x}^{\rm HF}-v_{\rm x}\right|\phi_\nu\right>-
\left<\phi_k\left|\hat{v}_{\rm x}^{\rm HF}-v_{\rm x}\right|\phi_k\right>
-\left<\psi_\nu \psi_k \left|g\right|\psi_\nu \psi_k\right>,\label{de1t}
\end{eqnarray}
and
\begin{eqnarray}
\Delta{\rm E}^{(1)}\left({\rm S},k\rightarrow\nu\right)=
\left<\phi_\nu\left|\hat{v}_{\rm x}^{\rm HF}-v_{\rm x}\right|\phi_\nu\right>-
\left<\phi_k\left|\hat{v}_{\rm x}^{\rm HF}-v_{\rm x}\right|\phi_k\right>
-\left<\psi_\nu \psi_k \left|g\right|\psi_\nu \psi_k\right>
+2\left<\psi_\nu \psi_k\left|g\right|\psi_k\psi_\nu\right>,\label{de1s}
\end{eqnarray}
where $g=1/\left|{\bf r}-{\bf r}'\right|$ and $\psi$ is just the spatial
component of the orbital $\phi$. $\hat{v}_{\rm x}^{\rm HF}$ is the Hartree-Fock
exchange potential, but constructed from the Kohn-Sham orbitals occupied in ground state
$\Phi_0$.

To evaluate the correction within the standard perturbation theory
(Eq.~\ref{H1}), we simply replace $v_{\rm x}$ with $v_{\rm xc}$ in the above
formulae.

\section{Results and discussion}
\label{s4}

In order to evaluate the excitation energies to first-order in the coupling
constant (Eqs.~\ref{de0},~\ref{de1t} and~\ref{de1s}) and in the more standard
perturbation scheme, we need to determine an accurate exchange-correlation
potential, Kohn-Sham orbitals and eigenvalues and the exchange component of the
exchange-correlation potential for the system of interest.

As mentioned in Sec.~\ref{s1}, the accurate exchange-correlation potential
used in Eq.~\ref{KS} for the determination of the Kohn-Sham orbitals and
eigenvalues is obtained starting from an accurate density and searching for
a potential that yields the desired density as a solution of the Kohn-Sham
equations (Eq.~\ref{KS} and~\ref{rho}).
In the special case of the singlet ground state of a two-electron
system~\cite{HE,harmonium}, the exchange-correlation potential can be simply
obtained from Eq.~\ref{KS}.
For systems with more than two electrons, $v_{\rm xc}$ can be
determined by expanding it in a complete set of basis functions and varying
the expansion coefficients such that Eqs.~\ref{KS} and~\ref{rho} yield the
accurate density~\cite{proc,UG,review}.
The exchange component of the exchange-correlation potential is simply
given by the condition that it cancels the self-interaction term in the
Hartree potential in the case of two electrons in the singlet ground state.
For a many-electron system, the exchange potential is obtained as discussed in
Ref.~\cite{separation}.

\begin{figure}[htb]
\centering
\centerline{\epsfxsize=9 cm \epsfbox{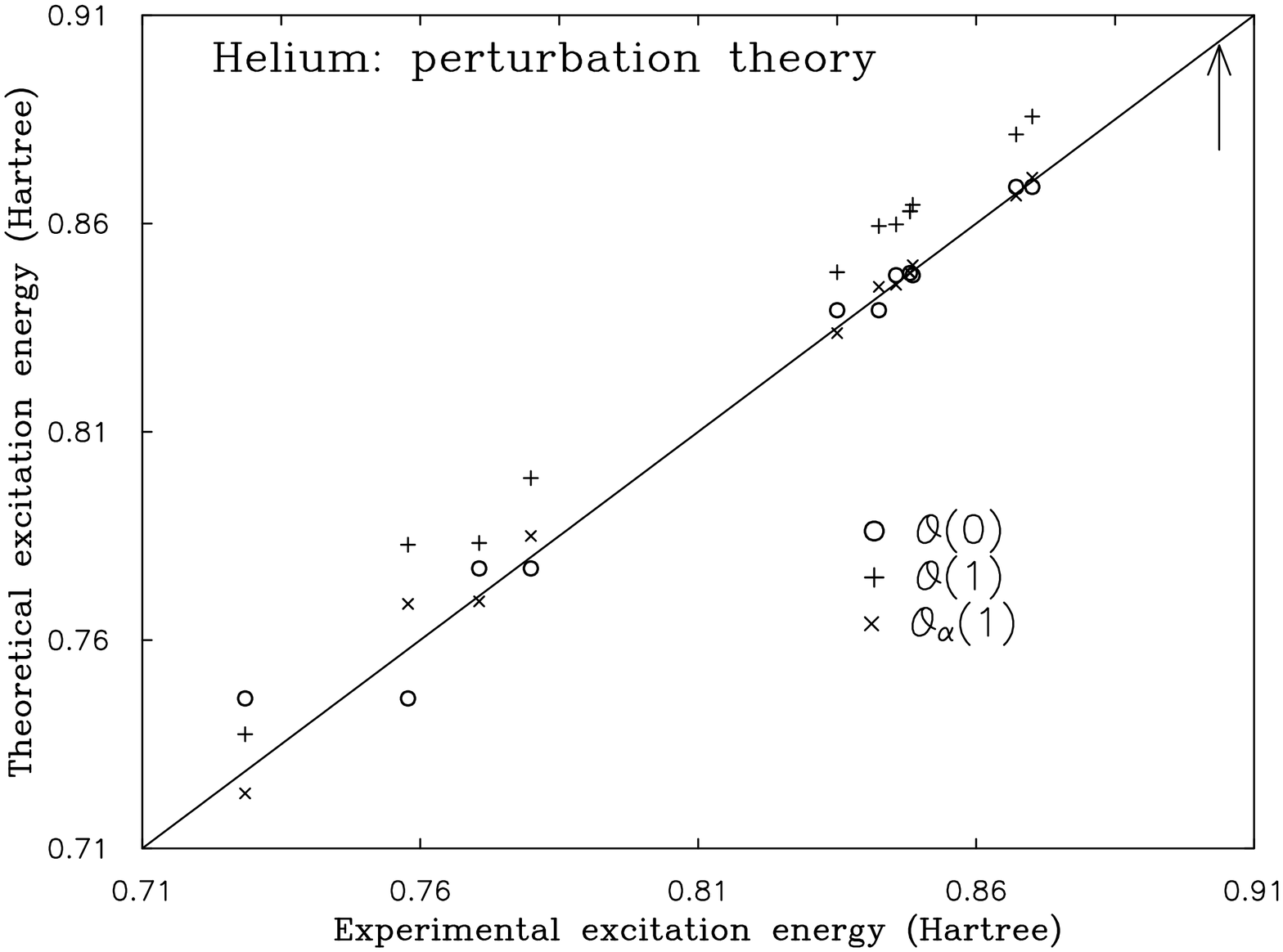}}
\vspace{.5cm}
\centerline{\epsfxsize=9 cm \epsfbox{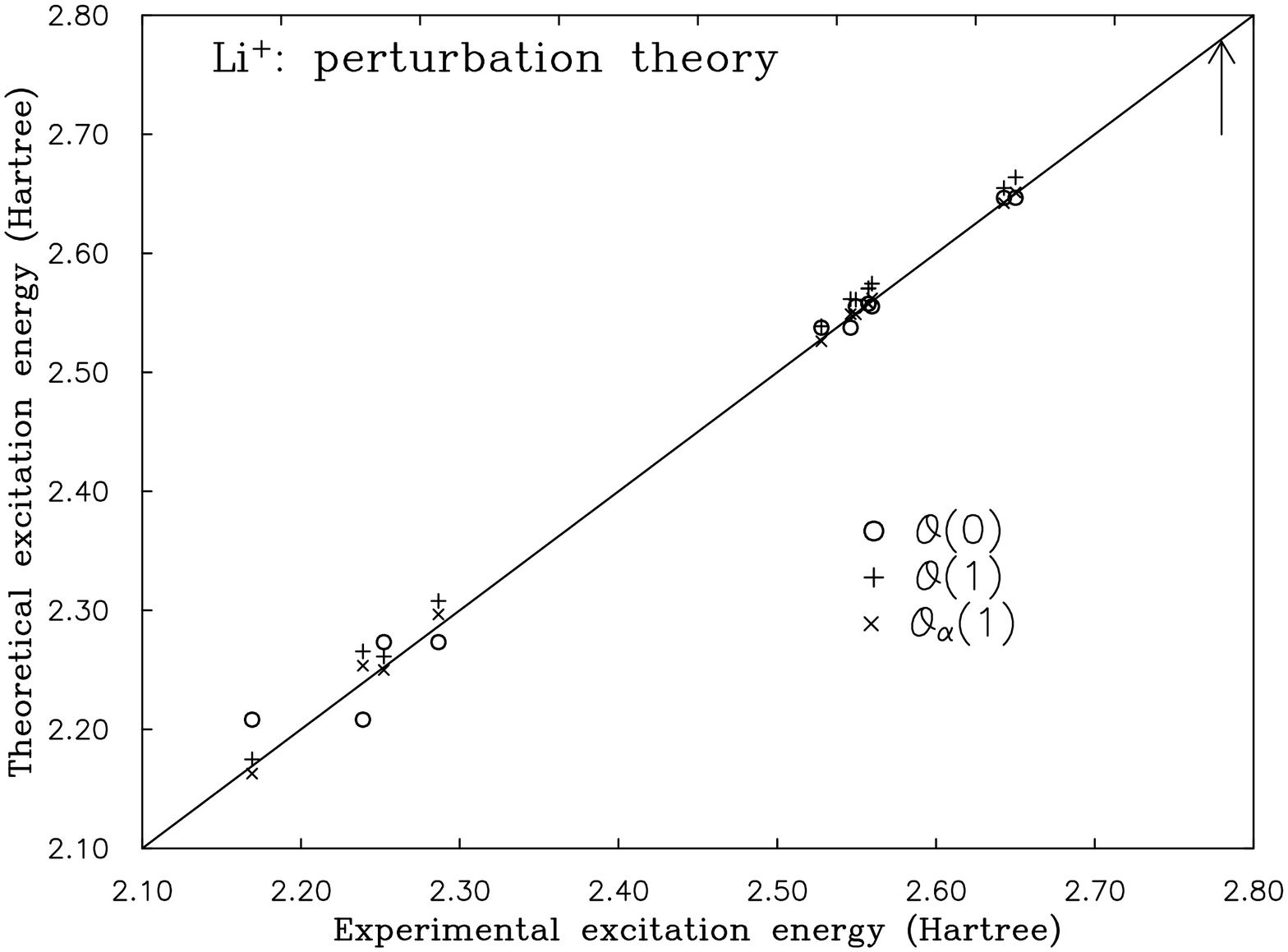}}
\vspace{.5cm}
\centerline{\epsfxsize=9 cm \epsfbox{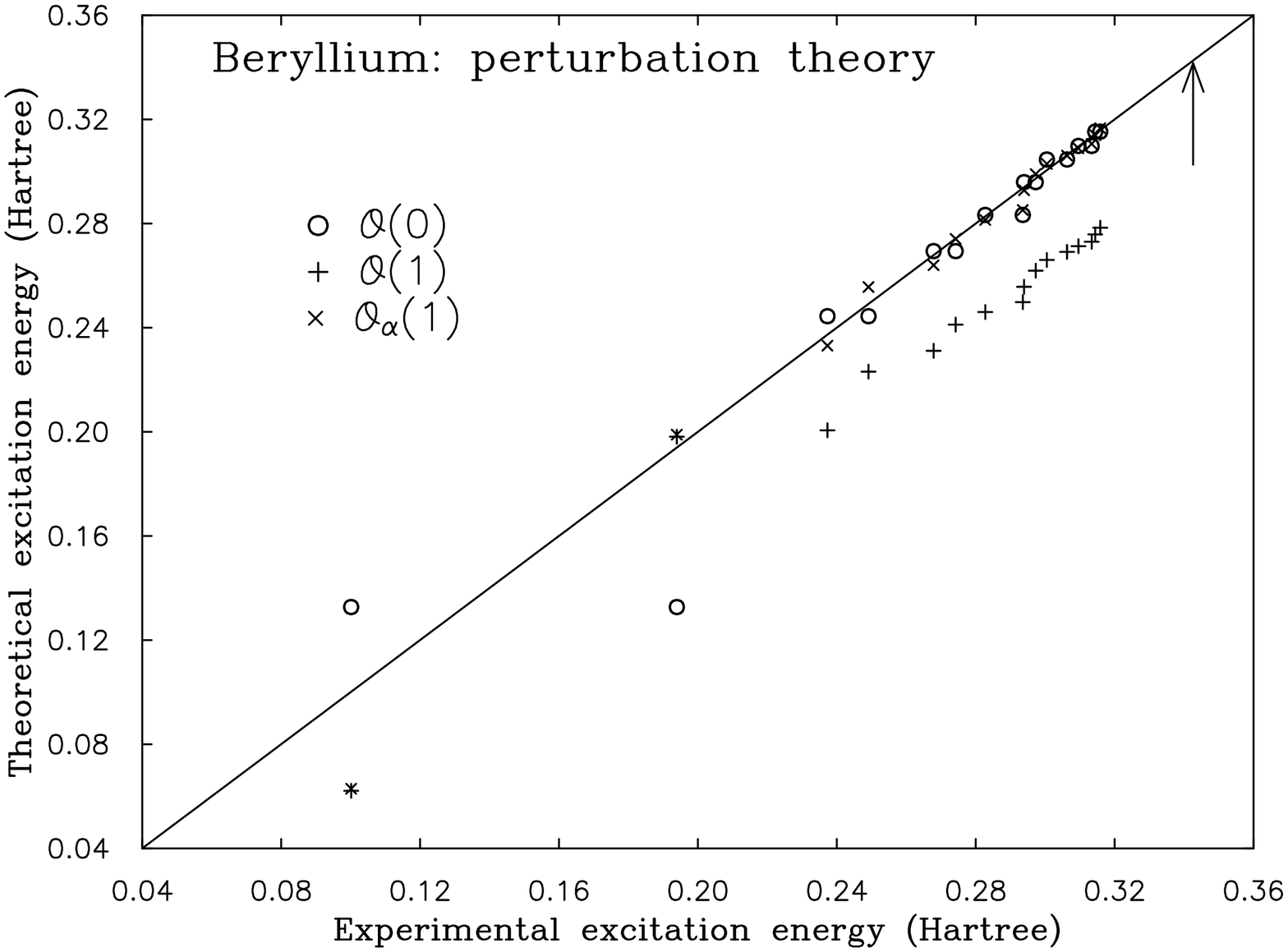}}
\vspace{.5cm}
\caption[]{Theoretical versus experimental excitation energies for He, Li$^+$
and Be (see Tables~\ref{table1}--\ref{table3}).
${\cal O}(0)$ is the zeroth-order perturbative result. ${\cal O}(1)$
and  ${\cal O}_\alpha(1)$ are the first-order results in the standard
and in coupling-constant perturbation schemes respectively.
The ionization limits are indicated by arrows.}
\label{f1}
\end{figure}

In Fig.~\ref{f1} and in Tables~\ref{table1},~\ref{table2} and~\ref{table3},
we show the excitation energies to zeroth-order and to first-order in the two
perturbative schemes for Helium, ionized Lithium and Beryllium respectively.

The zeroth-order excitation energies are simply given by the difference in
the Kohn-Sham eigenvalues and, as already discussed in Ref.~\cite{SUG},
lie between the experimental triplet and singlet excitation
energies to an accuracy of four significant digits, provided that
relativistic and finite nuclear mass effects are taken into account.
The only exception is the $1s\rightarrow 3d$ transition for ionized Lithium.

The first-order excitation energies obtained from standard perturbation theory
are on average less accurate than the zeroth-order energies. In contrast,
the excitation energies to first-order in the coupling-constant are on average
better than the zeroth-order energies but the improvement is not consistently
present for all the excitations.  With the exception of the $2s$ to $3d$,
$4p$ and $4d$ excitations for Beryllium, the singlet-triplet
splittings, obtained from either first-order scheme, are too large.
Note that for the two-electron systems, as shown in Appendix~\ref{a1},
Eqs.~\ref{de1t} and~\ref{de1s} yield a symmetrical
splitting around the zeroth-order excitation energy.

From Fig.~\ref{f1}, it is clear that the theoretical predictions for the
higher excitation energies converge to the experimental excitation energies
both in zeroth-order and in first-order coupling constant perturbation theory.
On the other hand, the excitation energies from first-order standard
perturbation theory converge to a too large value for Helium and ionized
Lithium and a too low value for Beryllium.
It is apparent that the zeroth-order energies must converge to the experimental
energies since the highest occupied Kohn-Sham eigenvalue is the negative of the
ionization energy~\cite{ioniz}.  In Fig.~\ref{f1}, the ionization limit for
each system has been indicated by an arrow.
There is an explanation of why the higher excitation energies from
first-order coupling constant perturbation theory converge to the
experimental excitation energies~\cite{Levy.pers}.
Coupling constant perturbation theory is based on the coupling constant
Hamiltonian in which the external potential is varied to keep the charge
density constant for all values of the coupling constant $\alpha$.
Since the ionization energy is related to the asymptotic decay of the
charge density, it is independent of the coupling constant.
Now, let us consider a perturbation expansion of the ionization energy and
charge density in powers of the coupling constant.
Since the ionization energy and density are independent of the coupling
constant, each term in the series, except the zeroth-order term, must be zero.
If we perform perturbation theory to a finite order, the ionization energy
is still strictly constant at each order: since the ionization energy is
obtained as an energy difference, each order of perturbation theory contributes
only to that same order.
We note in passing that this is not true of the density.  At a given order
of perturbation theory, the density is correct only to that order
since the $n$-th order wavefunction gives an $(n+1)$-th order contribution
to the charge density that is canceled only by going to the $(n+1)$-th order
of perturbation theory.
Finally, we note that the reason why the first-order energies in the standard
perturbation theory are too high for Helium and ionized Lithium and too low
for Beryllium is that the correlation potential, $v_{\rm c}$, is mostly positive
for the former and entirely negative for the latter~\cite{review}.
As explained in Sec.~\ref{s3}, the first-order excitation energies in
standard perturbation theory differ from the energies to first-order in the
coupling constant in having $v_{\rm xc}$ instead of only $v_{\rm c}$ in
Eqs.~\ref{de1t} and~\ref{de1s}.

We observe from Fig.~\ref{f1} that the zeroth-order and first-order in
coupling constant results are accurate not only close to the ionization
limit but over most of the energy range.  This calls for some further
explanation beyond that provided above.
In Ref.~\cite{SUG} the accuracy of the zeroth-order excitation energies
is attributed to the additional fact that the Kohn-Sham orbitals
and the quasi-particle amplitudes satisfy the same equation to order $1/r^4$.
%In this context we note that this argument applies only to excitations from
%the highest occupied orbital, in keeping with our finding that core excitations
%are not accurately predicted by zeroth or first-order perturbation theory.

We wish to emphasize that the accuracy of the first-order excitation energies,
obtained as differences of first-order total energies of the ground
and excited states, does not extend to the individual total energies of
the ground and excited states.
From Eq.~\ref{H1}, we obtain for the ground state energy
to first-order in the coupling constant:
\begin{eqnarray}
{\rm E}^{(1)}=
\langle \Phi_0 |\,{\cal H}^{(0)}+{\cal H}^{(1)}\,| \Phi_0\rangle =
\langle \Phi_0 |\,T + V_{\rm ext} +V_{\rm ee} +V_{\rm c}\,| \Phi_0\rangle.
\end{eqnarray}
Now, $\langle \Phi_0 |\,T + V_{\rm ext} +V_{\rm ee}\,| \Phi_0\rangle$
yields an energy that is a little worse than the energy from either the
optimized effective potential (OEP) method~\cite{OEP} or the Hartree-Fock method
since the Hartree-Fock wavefunction is the best single-determinant wavefunction,
while $\langle \Phi_0 |\,V_{\rm c}\,| \Phi_0 \rangle$ can be either negative or
positive.  For example, it is negative for Beryllium but positive for
Helium, ionized Lithium and Neon.  Consequently, at least for some systems,
the ground state energies obtained from the first-order formula are worse than
those from Hartree-Fock and the typical errors in Hartree-Fock total
energies is considerably larger than the errors we obtain for the
perturbative excitation energies.

We summarize the results in Table~\ref{rms} which shows the root mean square
errors in the excitation energies for our three test systems in zeroth-order
and first-order perturbation theory.
For the perturbation in the coupling-constant, the errors are reduced by about
a factor of two for Helium and Beryllium and a factor of three for ionized
Lithium.
For the standard perturbation theory, the errors are about a factor of two
worse than in zeroth-order for Helium and Beryllium while a slight
improvement is obtained for ionized Lithium.

In addition to the excitations from the highest occupied level, that we
have considered so far, we also studied the core-level excitation of
Beryllium from the 1s orbital to the 2p orbital,
1s$^2$ 2s$^2$ $^1$S $\to$ 1s$^1$ 2s$^2$ 2p$^1$ $^1$P.
The experimental energy is 4.243 Hartree~\cite{Becore}, while the
zeroth-order perturbative energy is 4.017 Hartree and the
first-order perturbative energies for the $^3$P and  $^1$P states
are 4.440 and 4.460 Hartree respectively.  Clearly, neither the zeroth-
order nor the first-order energies agree with experiment
as closely as do the energies of excitations from the highest occupied
level.
This is related to the fact that the eigenvalues of the occupied orbitals,
other than the highest, are not strictly related to the corresponding
ionization energies.

\begin{figure}[htb]
\centering
\centerline{\epsfxsize=9 cm \epsfbox{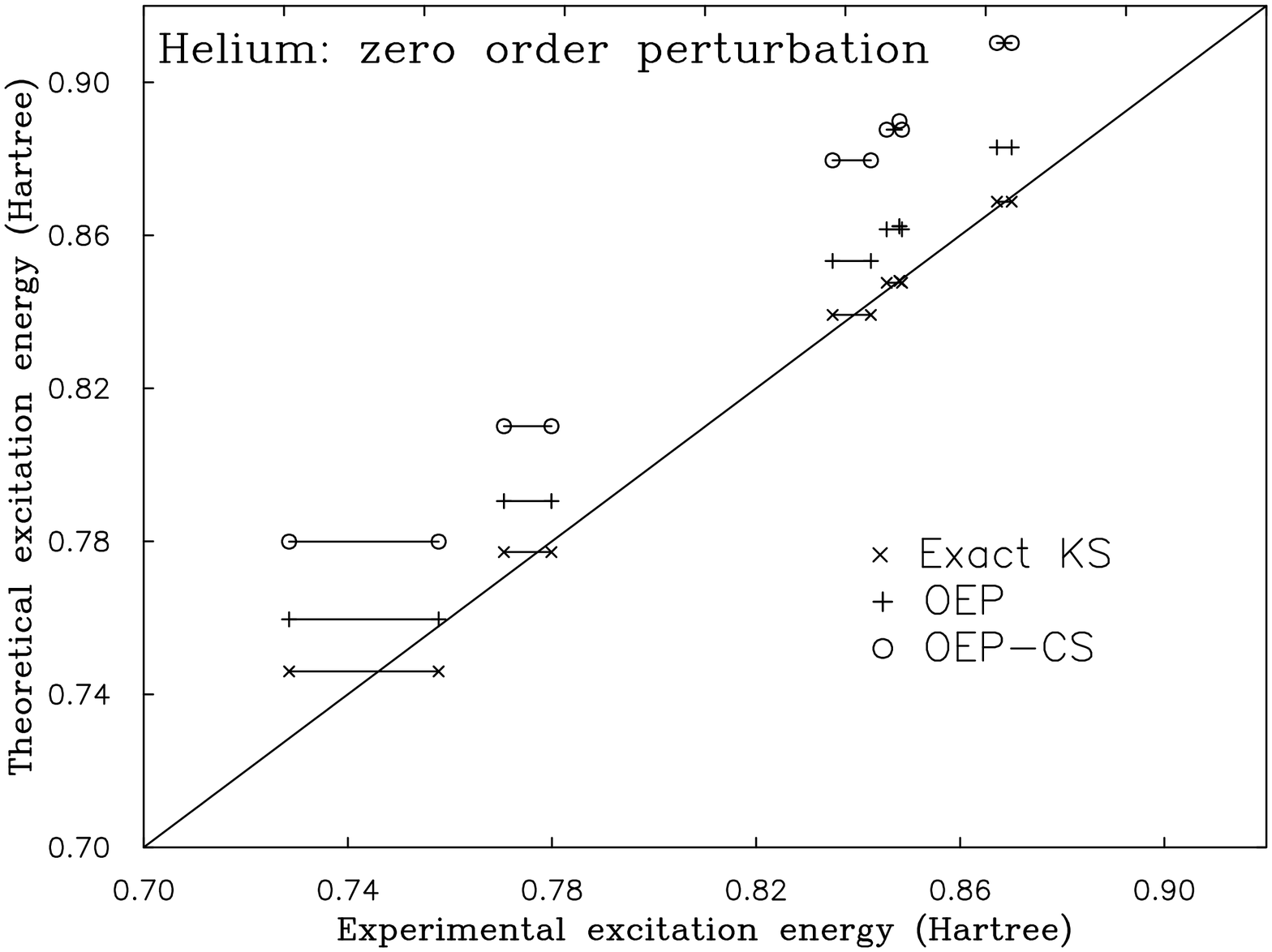}}
\vspace{.5cm}
\centerline{\epsfxsize=9 cm \epsfbox{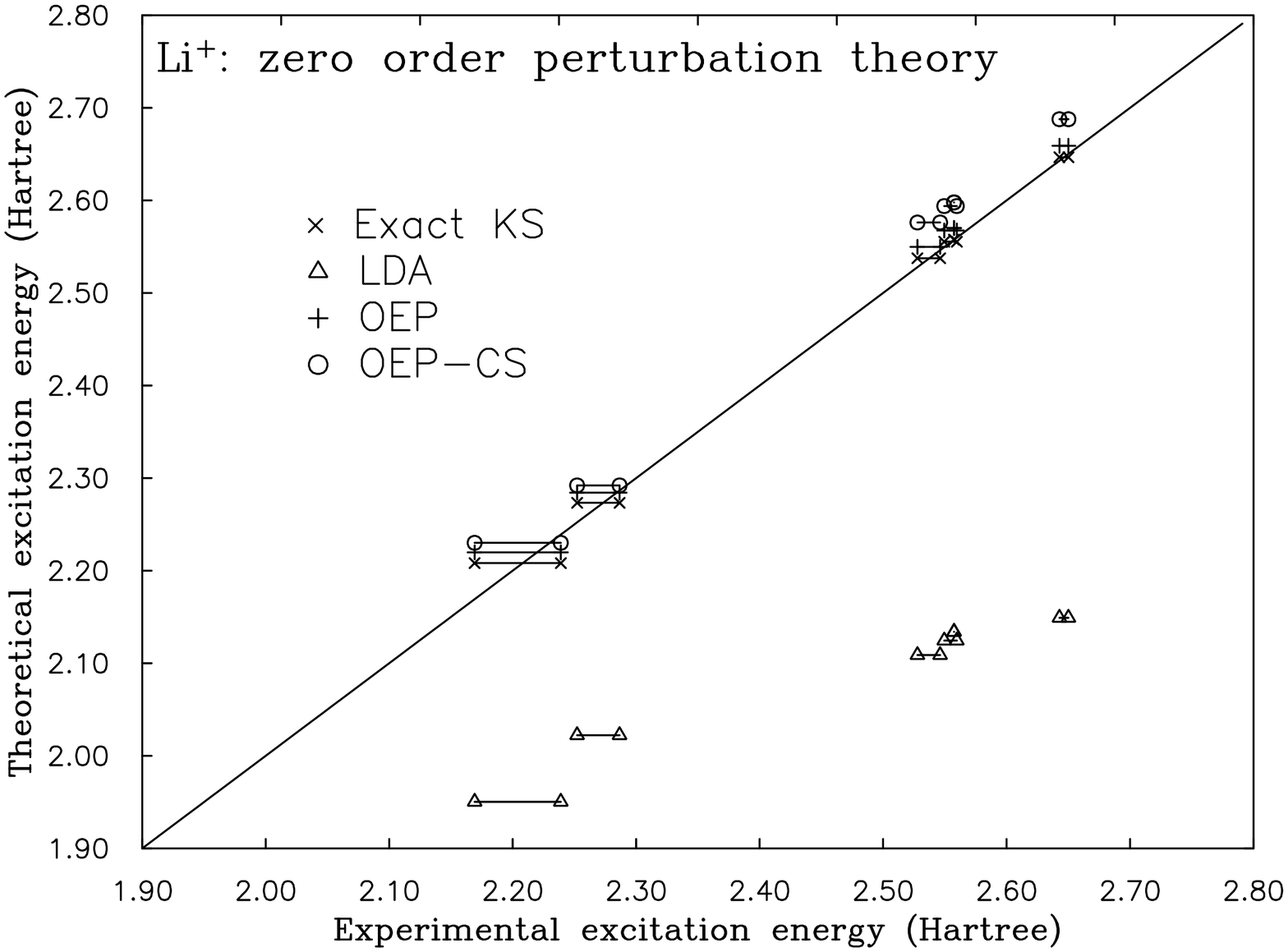}}
\vspace{.5cm}
\centerline{\epsfxsize=9 cm \epsfbox{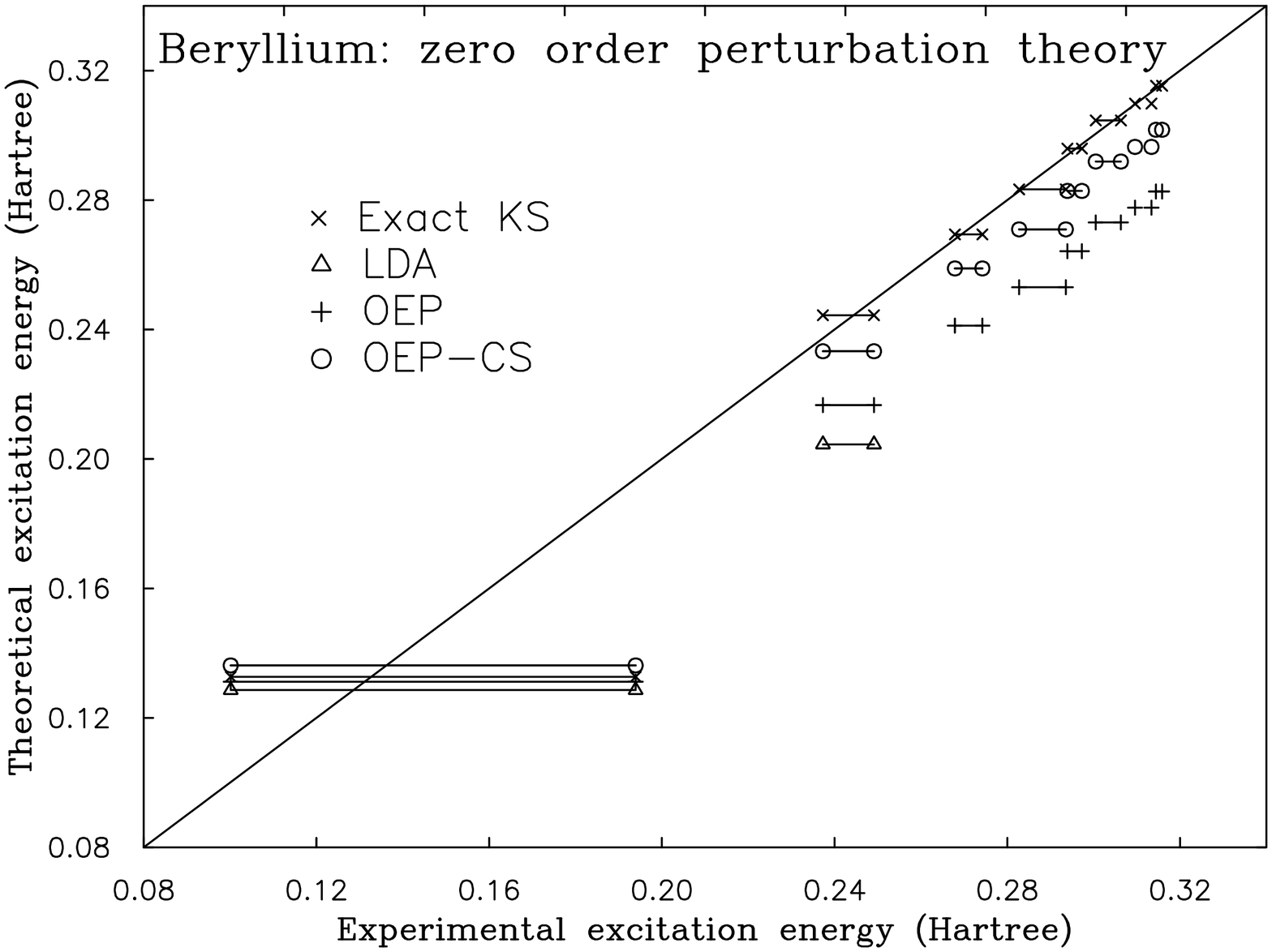}}
\vspace{.5cm}
\caption[]{Zeroth order perturbation theory excitation energies for He, Li$^+$
and Be. The theoretical excitation energies obtained from the exact Kohn-Sham,
LDA, OEP and OEP-Colle-Salvetti eigenvalues are plotted versus the
experimental values. The singlet and triplet energies from the same
configuration are joined by lines.}
\label{f2}
\end{figure}

Just as M\"oller-Plesset perturbation theory is commonly used in
quantum chemistry calculations, density functional functional perturbation
theory could become a feasible technique to access excited states.
However, the results presented above were obtained using the true Kohn-Sham
potentials whereas practical implementations of the method would need
to use potentials obtained from approximate exchange-correlation functionals.
In Fig~\ref{f2}, we compare the zeroth-order excitation energies
obtained from the local density approximation (LDA) and the optimized
effective potential method with those obtained from the true
Kohn-Sham potential.
The LDA does not bind any of the unoccupied orbitals in Helium and binds
only the lowest two unoccupied orbitals in Beryllium. Hence, the sparsity of
LDA data in Fig~\ref{f2}.
Since the highest occupied level in LDA is always too shallow, the
zeroth-order LDA excitation energies are on average too small.
However, note that the lowest excitation energies of Beryllium from all three
potentials are very close.  The reason is that this excitation is within the
same principal quantum number shell ($2s \to 2p$) and, consequently, both
orbitals involved in the excitation explore the same, rather limited,
spatial region.
The OEP excitation energies are too high for Helium and ionized Lithium and
too low for Beryllium, which of course is due to the correlation potential
being predominantly positive for the former and negative for the
latter~\cite{review}.

\begin{figure}[htb]
\centering
\centerline{\epsfxsize=9 cm \epsfbox{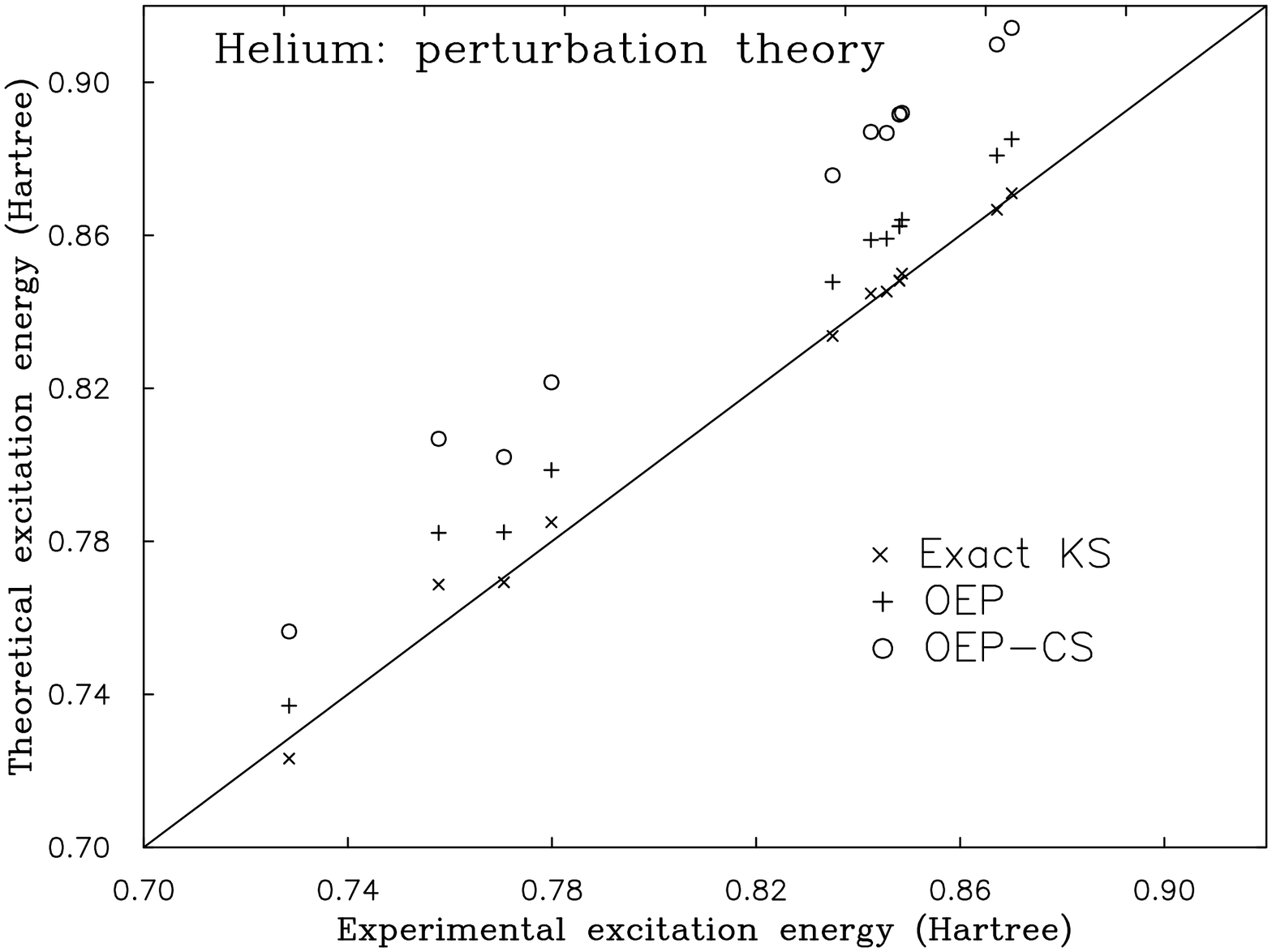}}
\vspace{.5cm}
\centerline{\epsfxsize=9 cm \epsfbox{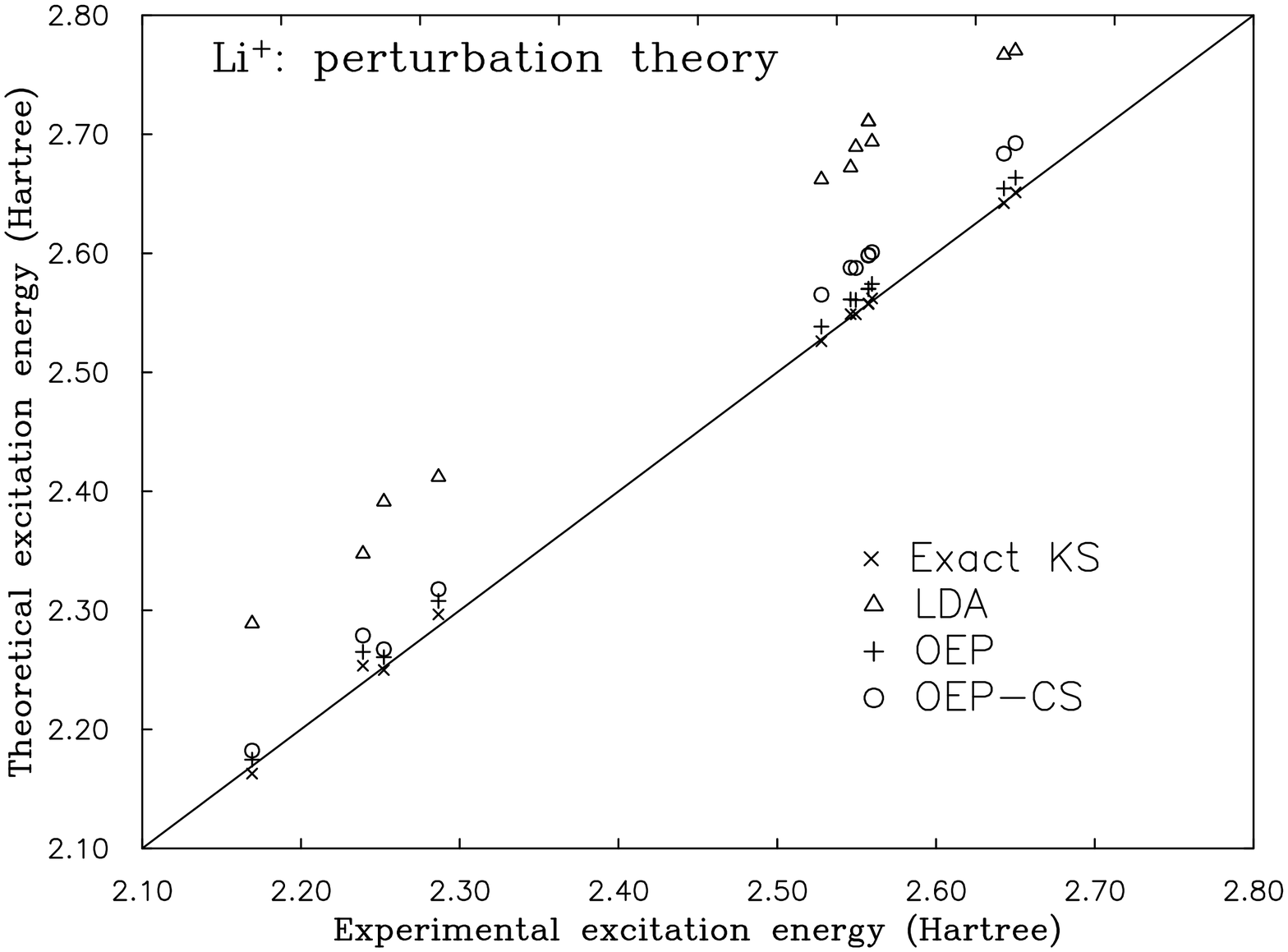}}
\vspace{.5cm}
\centerline{\epsfxsize=9 cm \epsfbox{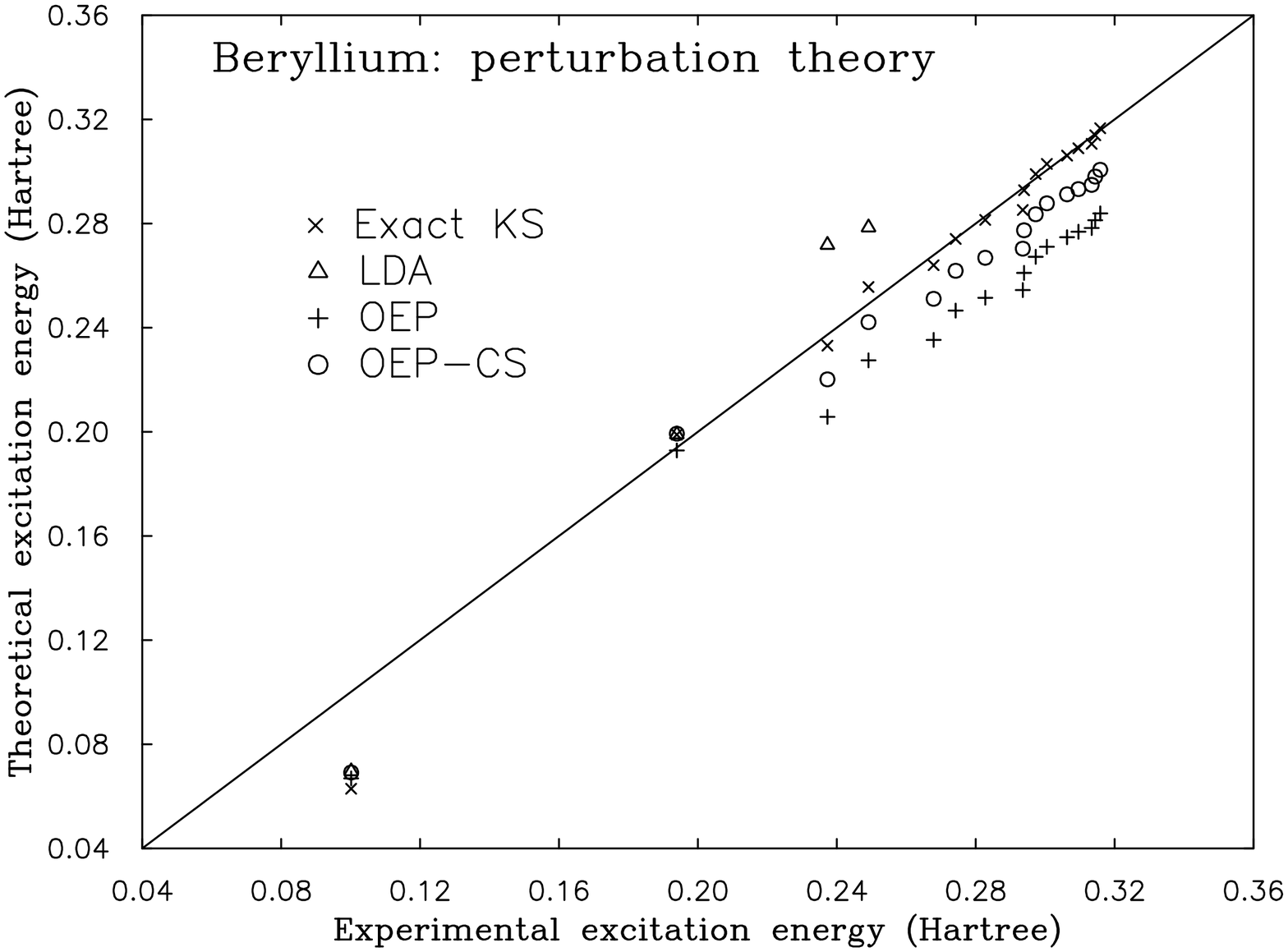}}
\vspace{.5cm}
\caption[]{Excitation energies for He, Li$^+$ and Be to first-order
perturbation theory in the coupling constant.
The theoretical excitation energies obtained from the exact Kohn-Sham
solutions, LDA, OEP and OEP-Colle-Salvetti are plotted versus the experimental
values.}
\label{f3}
\end{figure}

In Fig.~\ref{f3}, we present the same comparisons for the excitation
energies obtained within first-order coupling-constant perturbation theory.
The first-order OEP and first-order exact Kohn-Sham energies split
in a roughly symmetrical way about the corresponding zeroth-order
values while the first-order LDA energies have a large positive
shift relative to the LDA zeroth-order. The reason is
the following.  The second term in Eqs.~\ref{de1t} and~\ref{de1s},
$-\left<\phi_k\left|\hat{v}_{\rm x}^{\rm HF}-v_{\rm x}\right|\phi_k\right>$,
is strictly zero for both the true Kohn-Sham potential~\cite{LevyGorling96}
and for the OEP potential~\cite{KLI} when $\phi_k$ is the highest occupied
level.
On the other hand, this identity does not hold for the LDA exchange
potential which is in fact considerably shallower than the true one,
resulting in a large positive contribution from this term.
Note that all three quantities, $\phi_k$, $\hat{v}_{\rm x}^{\rm HF}$ and
$v_{\rm x}$, depend on whether they are being calculated for the true
Kohn-Sham theory, OEP or LDA but the main difference comes from the
change in $v_{\rm x}$.

We do not show GGA results in Figs.~\ref{f2} and \ref{f3} but we find
that the existing GGA's do not yield a large improvement upon the LDA
results. This is to be expected since the GGA potentials also fail to
have the correct asymptotic decay at large distances and are too shallow
for much of the region where the charge density is appreciable.

\begin{figure}[htb]
\centering
\centerline{\epsfxsize=9 cm \epsfbox{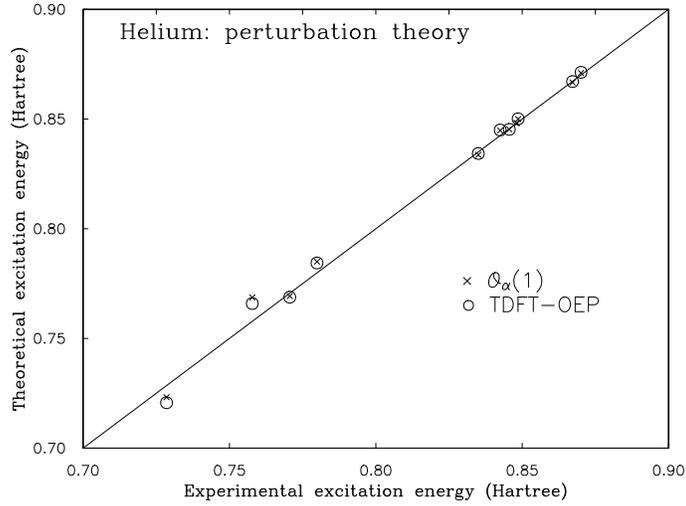}}
\vspace{.5cm}
\caption[]{Excitation energies for He within first-order perturbation
theory in the coupling constant and time-dependent density functional theory.
The time-dependent DFT excitation energies are obtained~\cite{PGG2} using the exact
Kohn-Sham potential as static potential and OEP for the dynamical response.
The theoretical excitation energies are plotted versus the experimental
values.}
\label{f4}
\end{figure}

Recently, there has been considerable interest in using time-dependent
density functional theory (TDFT) to calculate excitation energies.
In performing these calculations, a choice has to be made for the static
potential and for the dynamic response.
In Fig.~\ref{f4}, we compare the TDFT excitation energies of Helium, obtained
by Petersilka, Gossman and Gross~\cite{PGG2} using our exact Kohn-Sham
potential and the dynamic response from OEP, to our first-order
coupling-constant perturbation theory results.
The agreement is very close and on the scale of the plot
the difference between the two theories can only be detected for the
lowest two excitations.
For two electron systems, it is indeed possible to show that the TDFT
excitation energies obtained within the single-pole approximation,
using the exact Kohn-Sham potential and the time-dependent OEP
kernel, are strictly identical to the excitation
energies derived in first-order perturbation theory in the coupling constant
(Eqs.~\ref{de1t} and~\ref{de1s}). The fact that, in this case, the
single-pole approximation gives the leading correction to the Kohn-Sham
eigenvalue differences~\cite{PGG2} explains the close agreement found
in Fig.~\ref{f4}.

\begin{figure}[htb]
\centering
\centerline{\epsfxsize=9 cm \epsfbox{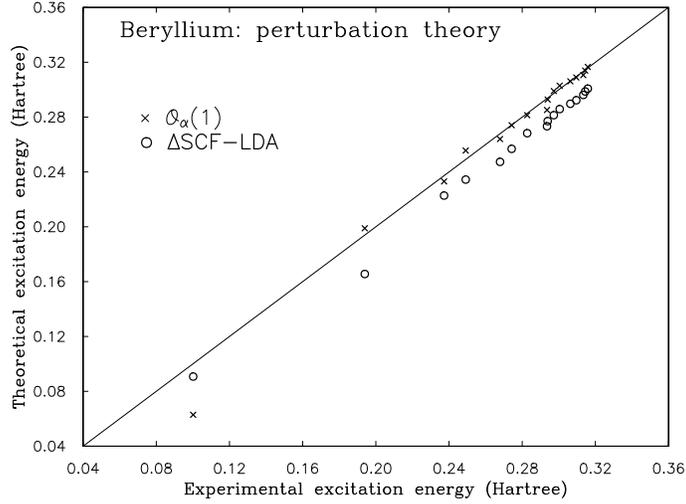}}
\vspace{.5cm}
\caption[]{Excitation energies for Be within first-order perturbation
theory in the coupling constant and $\Delta$SCF-LDA.
The theoretical excitation energies are plotted versus the experimental
values.}
\label{f5}
\end{figure}

As mentioned in the Introduction, an older method for calculating
excitation energies is the $\Delta$SCF method~\cite{VB,ZRB}.
Relative to perturbation theory, $\Delta$SCF has the advantage that,
when using approximate functionals, it can be applied even when the
corresponding approximate exchange-correlation potential does not bind
any unoccupied orbital.
%When using approximate functionals, this method has the advantage,
%relative to perturbation theory using the same approximate functional, that
%it can be applied even when the approximate exchange-correlation potential
%does not bound any unoccupied orbitals.
In Fig.~\ref{f5}, we compare the excitation energies of Beryllium from
$\Delta$SCF, using the LDA potential.
The results obtained for Beryllium are somewhat more accurate than those
from perturbation theory using the LDA or OEP potentials but of course
not as good as the perturbation theory results using the true Kohn-Sham
potential. Also for ionized Lithium, we find that, although less accurate than
the perturbation results with the exact Kohn-Sham solutions, the $\Delta$SCF
energies are more accurate than the LDA perturbative energies and of
comparable accuracy to the perturbative OEP results.

One can ask if it is possible to improve upon the perturbation theory
results obtained with the exact exchange (OEP) functional by
adding an approximate correlation functional.
Unfortunately, the correlation potentials obtained from the LDA and
various GGA correlation functionals bear no resemblance to the
true correlation potential~\cite{review}.
Recently, there has been considerable interest~\cite{GG} in orbital-dependent
correlation functionals, the most popular one being the Colle-Salvetti
functional~\cite{CS}.
In Figs.~\ref{f1} and \ref{f2}, we also show the correlation energies obtained
from adding the Colle-Salvetti correlation functional to the true exchange (OEP)
functional.  The energies are improved for Beryllium but are worse for Helium
and ionized Lithium since the Colle-Salvetti correlation potential
is mostly negative for all three systems whereas the true
correlation potential is negative for Beryllium but mostly positive for Helium
and ionized Lithium.

Clearly, a present obstacle to the success of perturbation theory
in the coupling constant seems to be the lack of an accurate approximation
for the exchange-correlation functional. Since approximate functionals such
as the local density or generalized gradient approximations bind few (if any)
unoccupied orbitals, combining an exact treatment of exchange with a local
orbital-dependent correlation functional, better than the Colle-Salvetti
functional,
appears to be the most promising possibility in this direction.

\begin{table}
\caption[]{Excitation energies of He in Hartree atomic units.
The theoretical energies of Drake~\cite{Drake93} and the perturbative
results are for infinite nuclear mass and neglect relativity.
${\cal O}(0)$ is the zeroth-order perturbative result. ${\cal O}(1)$
and  ${\cal O}_\alpha(1)$ are the first-order results in the standard
and in coupling-constant perturbation schemes respectively.}
\label{table1}
\begin{tabular}{llccccc}
Transition & Final State &  Experiment &   Drake & ${\cal O}(0)$ & ${\cal O}(1)$ & ${\cal O}_\alpha(1)$ \\[.1cm]
\hline\\[-.2cm]
           & $2\;^3S  $ & 0.7283   & 0.7285 & & 0.7374 & 0.7232 \\
$1s \rightarrow 2s$ & & & & 0.7460 & & \\
           & $2\;^1S  $ & 0.7576   & 0.7578 & & 0.7829 & 0.7687 \\[.1cm]
\hline\\[-.2cm]
           & $1\;^3P_o$ & 0.7704   & 0.7706 & & 0.7833 & 0.7693 \\
$1s \rightarrow 2p$ & & & & 0.7772 & & \\
           & $1\;^1P_o$ & 0.7797   & 0.7799 & & 0.7989 & 0.7850 \\[.1cm]
\hline\\[-.2cm]
           & $3\;^3S  $ & 0.8349   & 0.8350 & & 0.8483 & 0.8337 \\
$1s \rightarrow 3s$ & & & & 0.8392 & & \\
           & $3\;^1S  $ & 0.8423   & 0.8425 & & 0.8594 & 0.8448 \\[.1cm]
\hline\\[-.2cm]
           & $2\;^3P_o$ & 0.8455   & 0.8456 & & 0.8598 & 0.8453 \\
$1s \rightarrow 3p$ & & & & 0.8476 & & \\
           & $2\;^1P_o$ & 0.8484   & 0.8486 & & 0.8645 & 0.8500 \\[.1cm]
\hline\\[-.2cm]
           & $1\;^3D  $ & 0.8479   & 0.8481 & & 0.8629 & 0.8481 \\
$1s \rightarrow 3d$ & & & & 0.8481 & & \\
           & $1\;^1D  $ & 0.8479   & 0.8481 & & 0.8630 & 0.8482 \\[.1cm]
\hline\\[-.2cm]
           & $4\;^3S  $ & 0.8670   & 0.8672 & & 0.8814 & 0.8667 \\
$1s \rightarrow 4s$ & & & & 0.8688 & & \\
           & $4\;^1S  $ & 0.8700   & 0.8701 & & 0.8857 & 0.8710 \\
\end{tabular}
\end{table}

\begin{table}
\caption[]{Excitation energies of Li$^+$ in Hartree atomic units.
The theoretical energies of Drake~\cite{Drake2} and the perturbative
results are for infinite nuclear mass and neglect relativity.
${\cal O}(0)$ is the zeroth-order perturbative result. ${\cal O}(1)$
and  ${\cal O}_\alpha(1)$ are the first-order results in the standard
and in coupling-constant perturbation schemes respectively.}
\label{table2}
\begin{tabular}{llccccc}
Transition & Final State & Experiment & Drake & ${\cal O}(0)$ & ${\cal O}(1)$ & ${\cal O}_\alpha(1)$ \\[.1cm]
\hline\\[-.2cm]
         &  $2\;^3S $  & 2.1690 & 2.1692 &        & 2.1748 & 2.1628 \\
$1s \rightarrow 2s$ & &       & & 2.2082 & & \\
         &  $2\;^1S $  & 2.2330 & 2.2390 &        & 2.2655 & 2.2535 \\[.1cm]
\hline\\[-.2cm]
         &  $1\;^3P_o$ & 2.2521 & 2.2522 &        & 2.2612 & 2.2499 \\
$1s \rightarrow 2p$ & &       & & 2.2733 & & \\
         &  $1\;^1P_o$ & 2.2865 & 2.2866 &        & 2.3079 & 2.2966 \\[.1cm]
\hline\\[-.2cm]
         &  $3\;^3S $  & 2.5277 & 2.5278 &        & 2.5387 & 2.5261 \\
$1s \rightarrow 3s$ & &       & & 2.5375 & & \\
         &  $3\;^1S $  & 2.5460 & 2.5462 &        & 2.5615 & 2.5489 \\[.1cm]
\hline\\[-.2cm]
         &  $2\;^3P_o$ & 2.5493 & 2.5495 &        & 2.5611 & 2.5487 \\
$1s \rightarrow 3p$ & &       & & 2.5554 & & \\
         &  $2\;^1P_o$ & 2.5595 & 2.5597 &        & 2.5745 & 2.5621 \\[.1cm]
\hline\\[-.2cm]
         &  $1\;^3D $  & 2.5572 & 2.5574 &        & 2.5702 & 2.5574 \\
$1s \rightarrow 3d$ & &       & & 2.5576 & & \\
         &  $1\;^1D $  & 2.5574 & 2.5575 &        & 2.5706 & 2.5578 \\[.1cm]
\hline\\[-.2cm]
         &  $4\;^3S $  & 2.6426 & 2.6428 &        & 2.6548 & 2.6421 \\
$1s \rightarrow 4s$ & &       & & 2.6466 & & \\
         &  $4\;^1S $  & 2.6499 & 2.6501 &        & 2.6638 & 2.6511 \\[.1cm]
\end{tabular}
\end{table}

\begin{table}
\caption[]{Excitation energies of Be in Hartree atomic units.
The perturbative results are for infinite nuclear mass and neglect relativity.
${\cal O}(0)$ is the zeroth-order perturbative result. ${\cal O}(1)$
and  ${\cal O}_\alpha(1)$ are the first-order results in the standard
and in coupling-constant perturbation schemes respectively.}
\label{table3}
\begin{tabular}{llcccc}
Transition & Final State & Experiment & ${\cal O}(0)$ & ${\cal O}(1)$ & ${\cal O}_\alpha(1)$ \\[.1cm]
\hline\\[-.2cm]
         &  $1\;^3P_o$ & 0.1002 &         & 0.0621 & 0.0629 \\
$2s \rightarrow 2p$ & &        & 0.1327 & & \\
         &  $1\;^1P_o$ & 0.1939 &         & 0.1981 & 0.1989 \\[.1cm]
\hline\\[-.2cm]
         &  $2\;^3S $  & 0.2373 &         & 0.2006 & 0.2331 \\
$2s \rightarrow 3s$ & &        & 0.2444 & & \\
         &  $2\;^1S $  & 0.2491 &         & 0.2231 & 0.2556 \\[.1cm]
\hline\\[-.2cm]
         &  $2\;^3P_o$ & 0.2679 &         & 0.2311 & 0.2640 \\
$2s \rightarrow 3p$ & &        & 0.2694 & & \\
         &  $2\;^1P_o$ & 0.2742 &         & 0.2412 & 0.2741 \\[.1cm]
\hline\\[-.2cm]
         &  $1\;^3D $  & 0.2827 &         & 0.2460 & 0.2814 \\
$2s \rightarrow 3d$ & &        & 0.2833 & & \\
         &  $1\;^1D $  & 0.2936 &         & 0.2498 & 0.2852 \\[.1cm]
\hline\\[-.2cm]
         &  $3\;^3S $  & 0.2939 &         & 0.2557 & 0.2928 \\
$2s \rightarrow 4s$ & &        & 0.2959 & & \\
         &  $3\;^1S $  & 0.2973 &         & 0.2619 & 0.2990 \\[.1cm]
\hline\\[-.2cm]
         &  $3\;^3P_o$ & 0.3005 &         & 0.2660 & 0.3029 \\
$2s \rightarrow 4p$ & &        & 0.3046 & & \\
         &  $3\;^1P_o$ & 0.3063 &         & 0.2691 & 0.3061 \\[.1cm]
\hline\\[-.2cm]
         &  $2\;^3D $  & 0.3096 &       & 0.2713 & 0.3089 \\
$2s \rightarrow 4d$ & &        & 0.3098 & & \\
         &  $2\;^1D $  & 0.3134 &       & 0.2730 & 0.3106 \\[.1cm]
\hline\\[-.2cm]
         &  $4\;^3S $  & 0.3144 &       & 0.2758 & 0.3139 \\
$2s \rightarrow 5s$ & &        & 0.3153 & & \\
         &  $4\;^1S $  & 0.3156 &       & 0.2784 & 0.3166 \\[.1cm]
\end{tabular}
\end{table}

\begin{table}
\caption[]{Root mean square errors of the excitation energies of He, Li$^+$
and Be in zeroth-order and first-order perturbation theory.}
\label{rms}
\begin{tabular}{lccc}
System & ${\cal O}(0)$ & ${\cal O}(1)$ & ${\cal O}_\alpha(1)$ \\[.1cm]
\hline
He     & 0.0067 & 0.0159 & 0.0039 \\
Li$^+$ & 0.0167 & 0.0148 & 0.0056 \\
Be     & 0.0178 & 0.0358 & 0.0099 \\
\end{tabular}
\end{table}

\section*{Acknowledgements}
We thank Mel Levy, Andreas G\"orling and Andreas Savin for useful discussions
and critically reading the manuscript and Mel Levy for the explanation of
why the coupling-constant perturbation theory is more accurate than the
standard perturbation theory.
The calculations were performed on the IBM SP2 computer at the Cornell Theory
Center.
C.F. acknowledges financial support from NSF (grant number DMR 9422496).
C.J.U. is supported by the Office of Naval Research and X.G. by
FNRS-Belgium and the European Union (Human Capital and Mobility Program
contract CHRX-CT940462). C.J.U. and X.G. acknowledge support from
NATO (grant number CRG 940594).

\appendix
\section{First-order correction}
\label{a1}

We obtain here the expressions for the first-order energy differences given
in Eqs.~\ref{de1t} and~\ref{de1s}. We consider the ground state Kohn-Sham
determinant $\Phi_0$ and an excited determinant $\Phi_{k\nu}$ obtained
by exciting an electron from a doubly occupied orbital $k$ to an empty
orbital $\nu$.

First, we rewrite the contributions to the first-order energy difference
(Eq.~\ref{de1}) in terms of the Kohn-Sham orbitals:
\begin{eqnarray}
\left<\Phi_{k\nu}\left|v_{\rm H}+v_{\rm x}\right|\Phi_{k\nu}\right>-
\left<\Phi_0\left|v_{\rm H}+v_{\rm x}\right|\Phi_0\right>=
\left<\phi_\nu\left|v_{\rm H}+v_{\rm x}\right|\phi_\nu\right>-
\left<\phi_k\left|v_{\rm H}+v_{\rm x}\right|\phi_k\right>,\label{de.vxvh}
\end{eqnarray}
and
\begin{eqnarray}
\left<\Phi\left|V_{\rm ee}\right|\Phi\right>=
\frac{1}{2}\sum_{i,j}^{\rm occ}
\big(\left<ij\left|g\right|ij\right>-\left<ij\left|g\right|ji\right>\big),\nonumber
\end{eqnarray}
where $\left<ij\left|g\right|rs\right>\equiv\left<\phi_i\phi_j\left|g
\right|\phi_r\phi_s\right>=\int\int {\rm d}{\bf r}\,{\rm d}{\bf r}'\phi_i^*
({\bf r})\phi_j^*({\bf r}')\,g\,\phi_r({\bf r})\phi_s({\bf r}')$
and the sum is over the orbitals occupied in the determinant $\Phi$.
Since the Hartree potential $v_{\rm H}$
%and the Hartree-Fock exchange potential $\hat{v}_x^{\rm HF}$ are
is constructed from the
orbitals occupied in the ground state $\Phi_0$, we obtain
%(Eq.~\ref{hartree})
\begin{eqnarray}
\left<\Phi_{k\nu}\left|V_{\rm ee}\right|\Phi_{k\nu}\right>-
\left<\Phi_0\left|V_{\rm ee}\right|\Phi_0\right> &=&
\sum_{j}^{\rm occ}\big(
-\left<kj\left|g\right|kj\right>
+\left<kj\left|g\right|jk\right>
+\left<\nu j\left|g\right|\nu j\right>
-\left<\nu j\left|g\right|j\nu\right> \big)
-\left<\nu k \left|g\right|\nu k\right>
+\left<\nu k \left|g\right|k \nu\right>\nonumber\\
&=&
-\left<k\left|v_{\rm H}+\hat{v}_x^{\rm HF}\right|k\right>
+\left<\nu\left|v_{\rm H}+\hat{v}_x^{\rm HF}\right|\nu\right>
-\left<\nu k \left|g\right|\nu k\right>
+\left<\nu k \left|g\right|k \nu\right> \label{de.vee}
\end{eqnarray}
where the sums are over the Kohn-Sham orbitals occupied in $\Phi_0$
and $\hat{v}_x^{\rm HF}$ is the Hartree-Fock exchange potential constructed from
these orbitals.
Combining Eqs.~\ref{de1}, \ref{de.vxvh}, \ref{de.vee}, we obtain
%-esum_{j}^{\rm occ}\big(\left<kj\left|g\rightums are over the Kohn-Sham orbitals occupied in $\Phi_0$.
%jk\right>-
%\left<kj\left|g\right|kj\right>\big)
%-\sum_{j}^{\rm occ}\left<kj\left|g\right|kj\right>\nonumber\\
%&=&
%-\frac{1}{2}\sum_{i,j\neq k}^{\rm occ}\big(\left<ij\left|g\right|ji\right>
%-\left<ij\left|g\right|ij\right>\big)
%-\sum_{j}^{\rm occ}\left<kj\left|g\right|jk\right>\nonumber\\
%&=&
%-\frac{1}{2}\sum_{i,j\neq k}^{\rm occ}\big(\left<ij\left|g\right|ji\right>
%-\left<ij\left|g\right|ij\right>\big)
%-\left<k\left|\hat{v}_x^{\rm HF}\right|k\right>,\label{de.k}\\
%%
%\left<\Phi_{k\nu}\left|V_{\rm ee}\right|\Phi_{k\nu}\right>-
%\left<\phi_\nu\left|v_{\rm H}\right|\phi_\nu\right>&=&
%-\frac{1}{2}\sum_{i,j\neq k}^{\rm occ}\big(\left<ij\left|g\right|ji\right>
%-\left<ij\left|g\right|ij\right>\big)
%-\sum_{j\neq k}^{\rm occ}\big(\left< \nu j\left|g\right|j \nu \right>
%-\left< \nu j\left|g\right| \nu j\right>\big)
%-\sum_{j}^{\rm occ}\left< \nu j\left|g\right| \nu j\right>\nonumber\\
%&=&
%-\frac{1}{2}\sum_{i,j\neq k}^{\rm occ}\big(\left<ij\left|g\right|ji\right>
%-\left<ij\left|g\right|ij\right>\big)
%+\left<\nu\left|\hat{v}_x^{\rm HF}\right|\nu\right>
%+\left<\nu k\left|g\right|k\nu\right>-
%\left<\nu k \left|g\right|\nu k\right>,\label{de.nu}
%\end{eqnarray}
%where the sums are over the Kohn-Sham orbitals occupied in $\Phi_0$.
%Combining Eqs.~\ref{de1}, \ref{de.vxvh}, \ref{de.k} and \ref{de.nu}, we obtain
\begin{eqnarray}
\Delta{\rm E}^{(1)}=
\left<\phi_\nu\left|\hat{v}_x^{\rm HF}-v_{\rm x}\right|\phi_\nu\right>-
\left<\phi_k\left|\hat{v}_x^{\rm HF}-v_{\rm x}\right|\phi_k\right>
-\left<\phi_\nu \phi_k \left|g\right|\phi_\nu \phi_k\right>
+\left<\phi_\nu \phi_k \left|g\right|\phi_k \phi_\nu\right>. \label{de.a}
\end{eqnarray}

Our derivation of Eq.~\ref{de.a} is for a single-determinant $\Phi_{k\nu}$.
If the final state is a triplet, its Kohn-Sham wave function can be written
in terms of a single determinant provided that we consider either the $S_z=1$
or the $S_z=-1$ components of the triplet.  Then, the third term is zero
since we are considering the excitation from $\phi_k=\psi_k\chi_\downarrow$
to $\phi_\nu=\psi_\nu\chi_\uparrow$ or from $\phi_k=\psi_k\chi_\uparrow$ to
$\phi_\nu=\psi_\nu\chi_\downarrow$.
If $\Phi_{k\nu}$ is a singlet, it is necessary to write $\Phi_{k\nu}$
as a sum of two determinants. There are two
possible excitations, $\psi_k \chi_\uparrow\rightarrow\psi_\nu\chi_\uparrow$
and $\psi_k \chi_\downarrow\rightarrow\psi_\nu\chi_\downarrow$, and the
resulting wave function is equal to
$\Phi_{k\nu}=\left(\Phi_1-\Phi_2\right)/{\sqrt{2}}$.
Consequently,
\begin{eqnarray}
\left<\Phi_{k\nu}\left|V_{\rm ee}\right|\Phi_{k\nu}\right> =
\frac{1}{2}\left[\left<\Phi_1|V_{\rm ee}|\Phi_1\right>
+\left<\Phi_2|V_{\rm ee}|\Phi_2\right>
-\left<\Phi_1|V_{\rm ee}|\Phi_2\right>
-\left<\Phi_2|V_{\rm ee}|\Phi_1\right>\right].\nonumber
\end{eqnarray}
The first and second contribution are equal to each other and equivalent to
the above calculation~\ref{de.a}. The last two contributions are equal and, if we define
$\phi_1=\psi_k \chi_\downarrow$,
$\phi_2=\psi_\nu\chi_\uparrow$, $\phi_3=\psi_k \chi_\uparrow$ and
$\phi_4=\psi_\nu\chi_\downarrow$, give
\begin{eqnarray}
-\frac{1}{2}\left[\left<\Phi_1|V_{\rm ee}|\Phi_2\right>
+\left<\Phi_2|V_{\rm ee}|\Phi_1\right>\right]
=-\left<\phi_1\phi_2|g|\phi_3\phi_4\right>
+\left<\phi_1\phi_2|g|\phi_4\phi_3\right>\nonumber
\end{eqnarray}
The first term is zero due to spin orthogonality while the second term
can be rewritten as $\left<\psi_k\psi_\nu\left|g\right|\psi_\nu\psi_k\right>$.
Combining the above results, we obtain Eqs.~\ref{de1t} and~\ref{de1s}.

In Sec.~\ref{s3}, we mentioned that, for two-electrons in a singlet state,
formulae (Eq.~\ref{de1t} and~\ref{de1s}) yield a symmetrical splitting around
the zeroth-order excitation energy.
The ground state Kohn-Sham determinant of two electrons in a singlet state is
constructed from the two orbitals $\phi_1=\psi_0\chi_\uparrow$ and
$\phi_2=\psi_0\chi_\downarrow$. Here, we consider the excitation of an
electron from either $\phi_1$ or $\phi_2$ to the orbital $\phi_\nu$.

Since, for this system, the exchange potential satisfies
$v_{\rm x}=-v_{\rm H}/2=-\langle \psi_0 | g |\psi_0\rangle$, we obtain
\begin{eqnarray}
\langle \phi_1 |\, v_{\rm x}\,|\phi_1 \rangle =
\langle \phi_2 |\, v_{\rm x}\,|\phi_2 \rangle =
-\langle \psi_0 \psi_0 | g | \psi_0\psi_0 \rangle=
\langle \phi_1 |\,\hat{v}^{\rm HF}_{\rm x}\,| \phi_1 \rangle =
\langle \phi_2 |\,\hat{v}^{\rm HF}_{\rm x}\,|\phi_2 \rangle,\nonumber
\end{eqnarray}
so that
$\langle \phi_1 |\,\hat{v}^{\rm HF}_{\rm x}-v_{\rm x}\,| \phi_1 \rangle =
\langle \phi_2 |\,\hat{v}^{\rm HF}_{\rm x}-v_{\rm x}\,|\phi_2 \rangle=0$,
and
\begin{eqnarray}
\langle \phi_\nu |\,\hat{v}^{\rm HF}_{\rm x}-v_{\rm x}\,| \phi_\nu \rangle =
-\langle \psi_\nu\psi_0|g|\psi_0\psi_\nu\rangle+
\langle \psi_\nu\psi_0|g|\psi_\nu\psi_0\rangle.\nonumber
\end{eqnarray}
Consequently, by combining the above equations and the formulae (Eq.~\ref{de1t}
and~\ref{de1s}), we obtain a symmetrical splitting around the zeroth-order
excitation energy:
\begin{eqnarray}
\Delta{\rm E}^{(1)}\left({\rm T},0\rightarrow\nu\right)
=&-&\langle \psi_\nu\psi_0|g|\psi_0\psi_\nu\rangle,\nonumber\\
\Delta{\rm E}^{(1)}\left({\rm S},0\rightarrow\nu\right)
=& &\langle \psi_\nu\psi_0|g|\psi_0\psi_\nu\rangle.
\end{eqnarray}

\end{document}